\documentclass[twocolumn,showpacs,preprintnumbers,amsmath,amssymb]{revtex4}
\usepackage{graphicx}

\begin{document}

\draft

\title{NMR relaxation time around a vortex in stripe superconductors}

\author{M. Takigawa, M. Ichioka and K. Machida}
\address{Department of Physics, Okayama University,
         Okayama 700-8530, Japan}
\date{\today}

\begin{abstract}
Site-dependent NMR relaxation time $T_1({\bf r})$ is calculated 
in the vortex state using the Bogoliubov-de Gennes theory, 
taking account of possible "field-induced stripe'' states 
in which the magnetism arises locally 
around a vortex core in $d$-wave superconductivity.
The recently observed huge 
enhancement $T_1^{-1}({\bf r})$ below $T_c$ at a core site 
in Tl$_2$Ba$_2$CuO$_6$  is explained.
The field-induced stripe picture explains consistently 
other relevant STM and neutron experiments.
\end{abstract}

\maketitle

\narrowtext

Much attention has been focused on microscopic electronic structure, 
or low-lying excitation spectrum around a vortex core in a superconductor, 
particularly in the high $T_c$ cuprate superconductor, whose 
pairing mechanism is still not known.
It is expected that the vortex core might be a useful window through
which we can probe the ground state in the ''normal phase" by removing 
superconductivity.

Recently, keen interest is stimulated by a series of remarkable experiments:
Neutron scattering experiments observe an enhancement of the incommensurate 
magnetic satellite peaks by applying magnetic fields\cite{lake,katano,khaykovich}.
The checkerboard pattern of the local density of states (LDOS) in 
Bi$_2$Sr$_2$CaCu$_2$O$_{8+\delta}$, 
or the so-called "halo'' is found around a vortex site by scanning 
tunneling microscopy (STM) experiments by Hoffman {\it et al.}\cite{hoffman}.
Vortex-imaging NMR experiments detect anomalous relaxation time $T_1$ 
around a vortex core site in YBa$_2$Cu$_3$O$_7$\cite{halperin}
and YBa$_2$Cu$_4$O$_8$\cite{kakuyanagi}.
The set of these experiments collectively point to a notion 
that incipient spin density wave and/or charge density wave, 
namely, stripe states in a broad sense are induced around a vortex core by applying
a moderate magnetic field.

In $d$-wave pairing, which is now fairly well established for high $T_c$ cuprate 
superconductors, the low-lying vortex core excitations are expected to have
a finite density of states (DOS) exactly at the Fermi level.
Thus the core should be filled up by these quasi-particles and tunneling conductance 
as a function of the bias voltage  should exhibit a zero-energy peak (ZEP) structure
at a core. However, the existing STM results\cite{renner}
so far fail to detect this ZEP, instead they see the empty core, 
namely there is not any clear structure in the tunneling conductance at a core site.
This disparity between the theoretical prediction and experimental facts is 
rather serious, because in a conventional superconductor such as 2H-NbSe$_2$ 
the observed quasi-particle spectra by Hess {\it et al.}\cite{hess} 
are almost perfectly understood theoretically in terms of the bound 
state trajectories\cite{hayashi}. The situation becomes more serious 
since the vortex imaging NMR experiments done by Mitrovi\' c, 
{\it et al.}\cite{halperin} and Kakuyanagi, {\it et al.}\cite{kakuyanagi}
found that the site-dependent $T_1$ becomes longer at the vortex core, 
after being shorter with approaching towards the vortex. 
It is contradicted with existence of the ZEP of the LDOS, 
which should lead to the shortest $T_1$ at the core. 
This suggests absence of the ZEP associated with $d$-wave vortex, 
being rather consistent with the STM\cite{renner}.

The emerging notion of the field induced stripe matches with the above 
neutron experiments because the four site charge modulation corresponding 
to the STM observation\cite{hoffman} is accompanied by the eight site 
spin density wave, or stripe, seen as the incommensurate satellites. 
The applied field with a mere few Tesla enhances those satellite intensities 
and the estimated induced moment is roughly proportional to $H$\cite{lake}.

Prior to the present work, there have been several theoretical attempts 
to explain some of these facts, notably the STM experiments 
under $H$\cite{arovas,ogata,oka,demler,zhu,chen,martin,hedegard} and 
without $H$\cite{oka2,podolsky}. These theories are based on the field 
induced incipient spin and/or charge density orders, giving an 
explanation of the STM observation of four-site periodic LDOS modulation.

Quite recently, two new NMR imaging experiments\cite{kakuyanagi2,mitrovic} 
on the $T$-dependence of $T_1$ are reported: Kakuyanagi, 
{\it et al.}\cite{kakuyanagi2} show in Tl$_2$Ba$_2$CuO$_6$ that as lowering $T$, 
$1/T_1T$ at a core site increases divergingly towards a temperature 
$T_M$ below $T_c$ and then decreases at lower $T$.
This is contrasted with $1/T_1T$ at other sites which exhibit a monotonic 
decrease below $T_c$. This result suggests that below $T_M$ the magnetism 
appears locally exclusively at a vortex core site and other sites stay in 
the normally expected $d$-wave state.
Mitrovi\' c, {\it et al.}\cite{mitrovic} observe a similar divergent behavior 
in $1/T_1T$ of YBa$_2$Cu$_3$O$_7$ for the core site where the crossover 
temperature $T_M$ is quite low and $1/T_1T$ for the sites outside the core 
shows a constant at low $T$. The implication of these experiments is two-fold:
The missing ZEP in $d$-wave vortex core must be present 
in the limited $T$ and $H$ region. Below $T_M$ the locally field induced 
stripe must exist, which is to remove the  ZEP, giving rise 
to a suppression of $T_1^{-1}$ at lower $T$.

The purpose of this study is to demonstrate that the field induced stripe 
picture explains these experimental facts in a coherent way. 
This is done by explicitly calculating the site-dependent $T_1$, 
based on the Bogoliubov-de Gennes (BdG) equation for $d$-wave superconductors. 
Our calculation shows that the vortex imaging NMR\cite{takigawa,takigawa2,morr} 
yields a wealth of information on $d$-wave superconductors under $H$.

We begin with the standard Hubbard model on a two-dimensional
square lattice, and introduce the mean field
$ n_{i,\sigma}= \langle a^\dagger_{i,\sigma} a_{i,\sigma} \rangle$
at the $i$-site, where $\sigma$ is a spin index and $i=(i_x,i_y)$. 
We assume a pairing interaction $V$ between nearest-neighbor (NN) sites. 
This type of pairing interaction gives $d$-wave 
superconductivity\cite{takigawa,takigawa2,soininen}. 
Thus, the mean-field Hamiltonian under $H$ is given by 
\begin{eqnarray}
{\cal H}&=&
-\sum_{i,j,\sigma} \tilde{t}_{i,j}a^{\dagger}_{i,\sigma} a_{j,\sigma}
+U\sum_{i,\sigma} n_{i,-\sigma}
a^\dagger_{i,\sigma} a_{i,\sigma} 
\nonumber \\ &&
+V\sum_{\hat{e},i,\sigma}
(\Delta^\dagger_{\hat{e},i,\sigma} a_{i,-\sigma} a_{i+\hat{e},\sigma}
+\Delta_{\hat{e},i,\sigma} a^\dagger_{i,\sigma} a^\dagger_{i+\hat{e},-\sigma}
)\qquad 
\label{eq:2.2}
\end{eqnarray}
where $a^{\dagger}_{i,\sigma}$
($a_{i,\sigma}$) is a creation (annihilation) operator, and 
$i+\hat{e}$ represents the NN site ($\hat{e}=\pm\hat{x},\pm\hat{y}$).
The transfer integral is expressed as 
\begin{equation}
\tilde{t}_{i,j}=t_{i,j} \exp [ {\rm i}\frac{\pi}{\phi_0}\int_{{\bf
r}_i}^{{\bf r}_j} {\bf A}({\bf r}) \cdot {\rm d}{\bf r} ] , 
\label{eq:BdG4}
\end{equation}
with the vector potential 
${\bf A}({\bf r})=\frac{1}{2}{\bf H}\times{\bf r}$ in the symmetric gauge, 
and the flux quantum $\phi_0$.
For the NN pairs $(i,j)$, $t_{i,j}=t$.
For the next-NN pairs situated on a diagonal position on the square
lattice, $t_{i,j}=t'$. 
For the third-NN pairs, which are situated along the NN bond
direction, $t_{i,j}=t''$.
To reproduce the Fermi surface topology of cuprates, we set $t'=-0.12t$ and 
$t''=0.08t$\cite{tohyama}.  We consider the pairing interaction $V=2.0t$.
The essential results of this paper do not significantly depend on the 
choice of these parameter values. 

In terms of the eigen-energy $E_\alpha$ and the wave functions
$u_\alpha({\bf r}_i)$, $v_\alpha({\bf r}_i)$ at the $i$-site,
the BdG equation is given by
\begin{equation}
\sum_j
\left( \begin{array}{cc}
K_{\uparrow,i,j} & D_{i,j} \\ D^\dagger_{i,j} & -K^\ast_{\downarrow,i,j}
\end{array} \right)
\left( \begin{array}{c} u_\alpha({\bf r}_j) \\ v_\alpha({\bf r}_j)
\end{array}\right)
=E_\alpha
\left( \begin{array}{c} u_\alpha({\bf r}_i) \\ v_\alpha({\bf r}_i)
\end{array}\right) ,
\label{eq:BdG1}
\end{equation}
where
$K_{\sigma,i,j}=-\tilde{t}_{i,j} +\delta_{i,j} (U n_{i,-\sigma} -\mu)$, 
$D_{i,j}=V \sum_{\hat{e}}  \Delta_{i,j} \delta_{j,i+\hat{e}} $ 
and $\alpha$ is an index of the eigenstate\cite{takigawa,takigawa2}.   
The self-consistent condition for the pair potential 
and the number density is given by 
$\Delta_{i,j}
=\langle a_{j,\downarrow} a_{i,\uparrow} \rangle 
=\sum_\alpha  u_\alpha({\bf r}_i) v^\ast_\alpha({\bf r}_j) 
f(E_\alpha)$,
$n_{i,\uparrow}=\langle a^\dagger_{i,\uparrow} a_{i,\uparrow} \rangle 
=\sum_\alpha |u_\alpha({\bf r}_i)|^2 f(E_\alpha )$, 
$n_{i,\downarrow}=\langle a^\dagger_{i,\downarrow} a_{i,\downarrow}  \rangle 
=\sum_\alpha |v_\alpha({\bf r}_i)|^2 (1-f(E_\alpha )).$
The charge density $n_i=n_{i,\uparrow}+n_{i,\downarrow}$, 
the spin density $S_{z,i}=\frac{1}{2}(n_{i,\uparrow}-n_{i,\downarrow})$
and the staggered magnetization $M_{i}=S_{z,i}(-1)^{i_x+i_y}$. 
The $d$-wave order parameter at site $i$ is
$
\Delta_{d,i}=
( \Delta_{ \hat{x},i}
 +\Delta_{-\hat{x},i}
 -\Delta_{ \hat{y},i}
 -\Delta_{-\hat{y},i} )/4
$
with
\begin{equation}
\Delta_{\hat{e},i}={\bar \Delta}_{i,i + \hat{e}}
\exp[{\rm i}\frac{\pi}{\phi_0}
\int_{{\bf r}_i}^{({\bf r}_i+{\bf r}_{i + \hat{e}})/2}
{\bf A}({\bf r}) \cdot {\rm d}{\bf r}],
\nonumber
\label{eq:dOP2}
\end{equation}
where 
${\bar \Delta}_{i,i+\hat{e}}=\langle a_{i+\hat{e},\downarrow} 
a_{i,\uparrow} \rangle- \langle a_{i+\hat{e},\uparrow} 
a_{i,\downarrow} \rangle$.

We typically consider the case of a unit cell with $24 \times 24$ sites, 
where two vortices are accommodated.
The spatially averaged hole density is set to 
$n_{\rm h}=1-\overline{n_i} \sim \frac{1}{8}$ by tuning the 
chemical potential $\mu$. 
By introducing the quasimomentum of the magnetic Bloch state,
we obtain the wave function under the periodic boundary condition 
whose region covers many unit cells.

We construct Green's functions from $E_\alpha$, $u_\alpha({\bf
r})$, $v_\alpha({\bf r})$, and calculate the spin-spin correlation
function $\chi_{+,-}({\bf r},{\bf r}',i \Omega_n)$\cite{takigawa2}.
We obtain the nuclear spin relaxation rate,
\begin{eqnarray}
R({\bf r},{\bf r}') &=&
{\rm Im}\chi_{+,-}({\bf r},{\bf r}',
i \Omega_n \rightarrow \Omega + {\rm i}\eta)/(\Omega/T)|_{\Omega
\rightarrow 0}
\nonumber \\
&=&
 -\sum_{\alpha,\alpha'} 
[
 u_{\alpha}({\bf r})u^\ast_{\alpha}({\bf r}')
 v_{\alpha'}({\bf r})v^\ast_{\alpha'}({\bf r}')
\nonumber \\ &&
 -v_{\alpha}({\bf r})u^\ast_{\alpha}({\bf r}')
 u_{\alpha'}({\bf r})v^\ast_{\alpha'}({\bf r}')
 ]
\nonumber \\ &&
\times \pi T f'(E_\alpha) \delta(E_\alpha - E_{\alpha'})
\label{eq:T1}
\end{eqnarray}
with the Fermi distribution function $f(E)$.
We consider the case ${\bf r}={\bf r}'$ by assuming that
the nuclear relaxation occurs at a local site.
Then, ${\bf r}$-dependent relaxation time is given by $T_1({\bf
r})=1/R({\bf r},{\bf r})$.
We use $\delta(x)=\pi^{-1} {\rm Im}(x-{\rm
i}\eta)^{-1}$ to handle the discrete energy level of the finite
size calculation.
We typically use $\eta=0.01t$. In Eq. (\ref{eq:T1}), the first term is proportional to 
the $N_{\uparrow}({\bf r},E) \times N_{\downarrow}({\bf r},E)$ when ${\bf r}={\bf r}'$. 
To understand the behavior of $T_1({\bf r})$, we also consider the
LDOS given by $N({\bf r},E)
=N_{\uparrow}({\bf r},E)+N_{\downarrow}({\bf r},E)
=\sum_\alpha [|u_\alpha ({\bf r})|^2 \delta(E_\alpha -E)
+ |v_\alpha ({\bf r})|^2 \delta(E_\alpha +E)].$

\begin{figure}
\center{
\includegraphics[width=80mm]{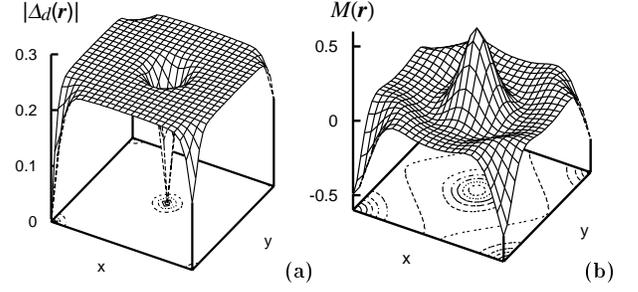}
}
\caption{\label{fig:1}
Spatial structures of the $d$-wave order parameter $\Delta_d({\bf r})$ and the
induced staggered moment $M({\bf r})$. 
At the vortex core site V, $|\Delta_{d}({\bf r})|=0$. 
The unit cell size is $24 \times 24$. 
$U/t=3.0$. $T/t=0.01$.
}
\end{figure}

By solving Eq.(\ref{eq:BdG1}) self-consistently, we obtain the
order parameter $\Delta_{d,i}$ and the sublattice magnetization $M_i$ as 
shown in Fig.1. It is seen that the moment is induced exclusively around 
the vortex core where the order parameter vanishes. 
The checkerboard modulation of $M$ with eight site period along 
$x$ and $y$ directions is superimposed. 
This is barely seen in Fig.1(b) where the ''floor'' in $M$ is modulated 
with eight-site period.
In the Fourier transformation of $S_{z,i}$ the eight-site period spin 
structure gives rise to peaks at the corresponding wave number ${\bf Q}$ 
in addition to the peak of the vortex lattice period.
As $U$ increases the checkerboard modulation 
with two ordering vectors ${\bf Q}=({3 \over 4}\pi,0)$ and 
$(0,{3 \over 4}\pi)$ changes 
into a one-dimensional stripe form characterized by 
${\bf Q}=({3 \over 4}\pi,0)$ or $(0, {3 \over 4}\pi)$ 
(see Fig.2 in Ref.\cite{oka} for stronger stripes). 
The induced moment at the core increases also with $U$.
This spin modulation accompanies the charge modulation with $2{\bf Q}$,
 namely, four site periodicity. 
The core accommodates excess particles, which is opposite
to the stripe-free case where it does excess holes\cite{hayashi2}.
These results basically coincide with the previous 
studies\cite{oka,demler,zhu,chen,martin} and also capture the essential 
features observed by STM experiment\cite{hoffman}.
The vortex core changes to a circular shape by induced moment as 
seen from Fig.1 (a). When $U=0$ its shape was square like.~\cite{hasegawa}

\begin{figure}
\center{
\includegraphics[width=60mm]{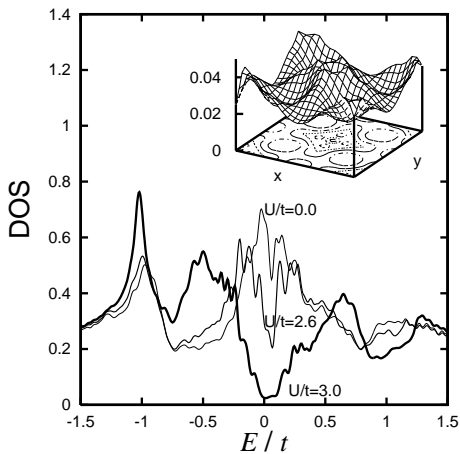}
}
\caption{\label{fig:2}
Local density of states $N({\bf r},E)$ at the core site for $U/t=0, 2.6, 3.0$.
The zero-energy state at $E=0$ for $U/t=0$ is gradually suppressed with $U$.
Inset shows the spatial profile of the  zero-energy LDOS $N({\bf r},E=0)$ 
for $U/t=3.0$. $T/t=0.01$.
}
\end{figure}

In Fig.2 we plot the LDOS as a function of the energy $E/t$ at the core site. 
For the small $U/t=2.6$ case where the appreciable moment is not induced, 
the ZEP appears, centered at $E/t=0$. As $U/t$  increases, 
the ZEP splits and is gradually suppressed. 
A gap-like feature becomes evident for $U/t=3.0$ in Fig.2. 
It is noted that the removed DOS at $E=0$ piles up inside the bulk gap at $|E/t|\sim 1$.
This DOS feature resembles the bound state at $E=$8meV observed  by 
STM on YBa$_2$Cu$_3$O$_7$\cite{renner}. 
The inset shows the spatial distribution of the zero-energy state: 
When the induced moment is absent ($U=0$), the zero-energy 
state accumulates at the core site. This is altered by introducing 
the induced moment, namely, $N({\bf r},E=0)$ 
decreases towards a core after taking a maximum. 
This is contrasted with the peak structure at a core 
in the stripe-free case (see Fig.6 in Ref.\cite{takigawa2}). 
We also note that the LDOS at $E=0$ shown in inset contains 
the Fourier component with 4-site checkerboard modulation.

\begin{figure}
\center{
\includegraphics[width=60mm]{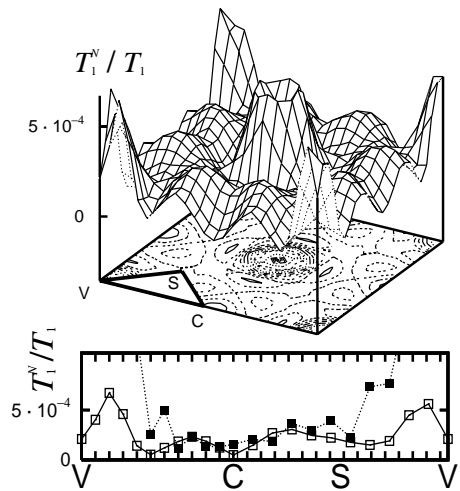}
}
\caption{\label{fig:3}
Spatial profile of $T_1^{-1}({\bf r})$ (upper panel).
$T_1^{-1}({\bf r})$ variation along the path (lower panel): 
V$\rightarrow$C$\rightarrow$S$\rightarrow$V
 where empty (filled) squares correspond to $U/t=3.0$ $(U/t=0)$.
 In the stripe-free case $(U/t=0)$ $T_1^{-1}({\bf r})$ at V is extremely large.
$T_1^N$ is the normal state value at $T_c$.
}
\end{figure}

We depict the spatial distribution for $T_1^{-1}({\bf r})$ 
at low $T(=0.01t)$ for $U/t=3.0$ in upper panel of Fig.3.
It is seen that corresponding to the depressed LDOS 
at the core shown in Fig.2 due to the induced moment, 
$T_1^{-1}({\bf r})$ diminishes 
around the vortex core after it increases 
with approaching the vortex from far sites.
In lower panel of Fig.3, 
$T_1^{-1}({\bf r})$ is depicted along the selected 
paths in the vortex lattice shown in upper panel. 
It is clearly seen that (1) $T_1^{-1}({\bf r})$ 
at the core site V takes a local minimum. 
(2) One of the minima of $T_1^{-1}({\bf r})$ occurs at C-point. 
(3) The relative $T_1^{-1}({\bf r})$ values at V and C points 
depend on the magnitude of the induced moment. 
When $|M|$ is large enough, $T_1^{-1}({\bf r})$ becomes 
the absolute minimum at V.
(4) The maximum $T_1^{-1}({\bf r})$ occurs near the core 
site whose distance from the core depends on $|M|$. 
In larger $|M|$ case this distance becomes large 
because of the larger LDOS suppression due to the induced moment.

Both site-selective experiments of $^{17}$O NMR 
on YBa$_2$Cu$_3$O$_7$\cite{halperin} and YBa$_2$Cu$_4$O$_8$\cite{kakuyanagi} 
for the CuO$_2$ planar site show that $T_1^{-1}({\bf r})$
at the core is smaller than the neighboring sites 
for a certain $H$ and $T$ region.
The $T_1^{-1}({\bf r})$ maximum occurs in the vicinity of the core, 
{\it i.e.} $T_1^{-1}({\bf r})$ decrease with increasing 
the internal field around the vortex core region.
These facts  are reproduced by our calculation, evidencing 
the usefulness of the site-selective NMR method in general.
When we see the $T_1^{-1}({\bf r})$ variation 
along C$\rightarrow$S$\rightarrow$V in the experimental 
data\cite{halperin,kakuyanagi}, $T_1^{-1}({\bf r})$ 
has a minimum near the saddle point S. This behavior is 
also reproduced in our calculation of the stripe state here,
while in the previous calculation ($U=0$), 
$T_1^{-1}({\bf r})$ monotonically increases along 
C$\rightarrow$S$\rightarrow$V as displayed 
in lower panel in Fig.3.

We show the $T$-dependence of $1/T_1({\bf r})T$ 
for three sites in Fig.4. 
It is seen that $1/T_1({\bf r})T$ at the vortex 
site V displays a large enhancement around $T_M$ 
below which $M$ begins to appear as shown in the inset. 
Thus for $T_M<T<T_c$ the zero-energy state increases 
$1/T_1T$ and for $T<T_M$ the induced moment opens 
a gap at $E=0$ as shown in Fig.2, reducing $1/T_1T$ 
towards low $T$.
The $1/T_1T$ behavior in the nearest neighbor site 
to the core is moderately enhanced at $T_M$
while that in S-site remains same as in the stripe-free case. 
These correspond to the fact that the induced moment is 
limited only near the core. We should remark here that 
the enhancement in $1/T_1T$ below  $T_c$ delicately 
depends on applied field. As $H$ increases, 
it is suppressed rather easily (see Fig.12 in Ref.\cite{takigawa2}).

\begin{figure}
\center{
\includegraphics[width=60mm]{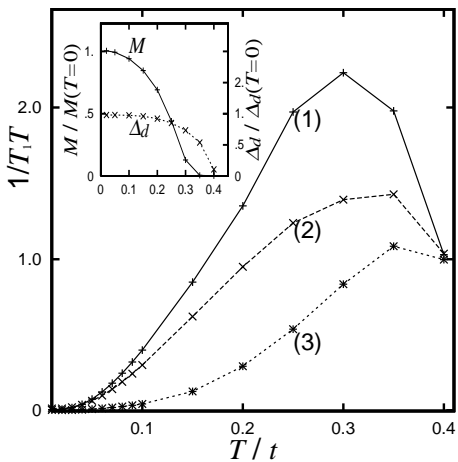}
}
\caption{\label{fig:4}
$T$-dependence of $1/T_1({\bf r})T$ for three sites. 
(1) The vortex core (V).
(2) The nearest site to the core. 
(3) The saddle point site (S). 
Inset shows the corresponding $T$-dependence 
of $\Delta_d$ and $M$ at their maximum values. $U/t=3.0.$
}
\end{figure}

The diverging behavior in $1/T_1({\bf r})T$ at the core site is 
strikingly similar to that observed in $^{205}$Tl NMR on 
Tl$_2$Ba$_2$CuO$_6$\cite{kakuyanagi2}. Our calculation is 
also consistent with $1/T_1({\bf r})T$ behavior at the S point:
Their data show strong decrease of $1/T_1({\bf r})T$ at S below $T_c$ and 
$1/T_1T$=constant at lower $T$, indicative of the zero-energy state.

As for Mitrovi{\'c}'s data\cite{mitrovic}, they observe that 
(A) $1/T_1T$ inside the core is enhanced as lowering $T$
while (B) $1/T_1T$ outside the core stays at a constant 
which increases as $H$ increases. These behaviors are understandable 
if we assume that $T_M$ is low in YBa$_2$Cu$_3$O$_7$, namely
their experiments are done for $T>T_M$:
We can explain the fact (A) because as lowering $T$ the thermal 
smearing factor coming from $f'(E)$ in eq.(\ref{eq:T1}) 
becomes narrow enough to see the sharp ZEP, 
driving $1/T_1T$ to increase
towards $T\rightarrow 0$. The fact (B) is consistent with 
the previous more detailed NMR at the saddle point by Zheng, 
{\it et al.}\cite{zheng} on YBa$_2$Cu$_4$O$_8$: They conclude that 
$1/T_1T\propto H$ at the saddle point at low $T$.  
As shown previously the spatial average DOS 
$N(0)\propto H^{0.41}$ for $d$-wave case\cite{hasegawa}, 
then $1/T_1T\propto N(0)^2\sim H^{0.82}$, explaining 
the approximate $H$-linear behavior within the present 
experimental accuracy.

Many ordinary $T_1$ measurements are done by using 
the resonance field at the maximum intensity which 
corresponds to the saddle point S of the field distribution
in the vortex state. Above $T_c$, $1/T_1T$ shows a variety 
of different $T$-behaviors, depending on probed nuclei, 
such as O, Cu, Y, etc, or materials, or doping levels.
These concern the so-called pseudo-gap phenomena. In contrast,
the low $T$ behaviors, remarkably enough, are quite similar, 
namely, a $T^3$-law below $T_c$ which is crossed over to 
the $T$-linear law at lowest $T$. The former $T^3$-law 
indicates a line node for $d$-wave pair while the $T$-linear 
law is indicative of the zero-energy state outside of the vortex core, 
in spite of possible existence of the induced stripe. 
This is due to the fact that the induced moment is confined 
in the narrow core region and LDOS or $T_1^{-1}$ is modified 
only there as seen from the lower panel of Fig.3.

In conclusion, we have calculated the site-dependent relaxation time 
$T_1^{-1}({\bf r})$ for NMR, taking into account the possibility of 
the field-induced moment. The large $T_1^{-1}({\bf r})$ enhancement 
observed in the vortex core site NMR in $^{205}$Tl of 
Tl$_2$Ba$_2$CuO$_6$ is explained by our calculation. 
Other experiments for the vortex imaging\cite{halperin,kakuyanagi,mitrovic} are
also analyzed in a coherent manner. 
An emerging notion through these analyses, 
together with other STM and neutron scattering experiments, 
is the field-induced stripe picture in the vortex state of 
high $T_c$ superconductors. The strength of the induced stripe 
depends on material's parameters. The existing data show 
that Tl$_2$Ba$_2$CuO$_6$ is strong while YBa$_2$Cu$_3$O$_7$ 
and YBa$_2$Cu$_4$O$_8$ are weak.

\noindent
We thank K. Kumagai, Y. Matsuda and K. Ishida for useful discussions on their NMR.

\end{document}